\documentclass[english,reprint,aps,prl,superscriptaddress,nofootinbib]{revtex4-1}
\usepackage[T1]{fontenc}
\usepackage[latin9]{inputenc}
\usepackage{geometry}
\usepackage{color}
\usepackage{xcolor}
\usepackage{xr-hyper}
\usepackage{hyperref}
\usepackage{amsmath}
\usepackage{amssymb}
\usepackage{stmaryrd}
\usepackage{graphicx}

\makeatletter

\usepackage{babel}

\makeatother

\usepackage{babel}
\begin{document}

\title{Dark Andreev States in Superconductors}

\author{Andrey Grankin and Victor Galitski}
\affiliation{Joint Quantum Institute, Department of Physics, University of Maryland, College Park, MD
20742, USA}

\begin{abstract}
The conventional Bardeen-Cooper-Schrieffer (BCS) model of superconductivity assumes  a frequency-independent order parameter, which allows a relatively simple description of the superconducting state. In particular, its excitation  spectrum readily follows from the Bogoliubov-de-Gennes (BdG) equations. A more realistic description of a superconductor is the Migdal-Eliashberg theory, where the pairing interaction, the order parameter, and electronic self-energy are strongly frequency dependent. This work combines these ingredients of phonon-mediated superconductivity with the standard BdG approach. Surprisingly, we find qualitatively new features such as  the emergence of a shadow superconducting gap in the quasiparticle spectrum at energies close to the Debye energy. We show how these features reveal themselves in standard tunneling experiments. Finally, we also predict the existence of additional high-energy bound states, which we dub ``dark Andreev states.'' 
\end{abstract}

\maketitle

Bardeen-Cooper-Schrieffer theory of superconductivity \citep{BCS57}
and Bogoliubov-de Gennes equations have proven to be relatively simple
and reliable tools to describe a variety of conventional superconductors.
A key simplifying assumption of this approach is that the superconducting
order parameter is energy-independent~\citep{AS10}. While it is
clearly not the case in any real superconductor, many features such
as the quasiparticle spectrum, thermodynamic and electromagnetic properties
\citep{AS10, BCS57} appear insensitive to this approximation. It
is reasonable, because the omitted energy dependence normally does
not affect low-energy physics. In contrast, accurate determination
of the transition temperature is sensitive to the details of the phonon
dispersion, the dynamical screening of Coulomb interaction, and the
structure of the superconducting gap. The latter can be obtained from
the Migdal-Eliashberg (ME) equations \citep{E60, M20} which are integral
equations in both energy and momentum.

This Migdal-Eliashberg theory~\cite{MSC88} predicts a non-trivial
structure of the superconducting gap as a function of {\em real}
frequency. In particular, it was shown that for superconductivity
mediated by the Einstein phonon modes~\citep{M18,MBM20}, the gap
function has sharp resonance-like features. In the weak coupling regime,
these features can be well approximated by the Lorentz function centered
at the Debye frequency. Another interesting real-frequency behavior
of the gap function is discussed in \citep{CC21} where the authors
predict the possibility of formation of frequency-domain vortices
in the presence of phonon-induced attraction and Coulomb repulsion.

In this work, we study how a frequency-dependent order parameter affects
the quasiparticle spectrum and tunneling properties of a superconductor.
We assume that the frequency dependence has a Lorentz shape, which
corresponds to optical-phonon-mediated pairing~\citep{M20}. We solve the corresponding
generalized BdG equations and show that an additional gap emerges
in the qusiparticle spectrum at higher energies. In the case of an
SNS junction, we also find that additional Andreev in-gap high-energy
states can form. We use analogy with quantum optics to interpret these
high-energy peaks in terms of ``dark'' resonance features~\citep{FL00,LYF99,FM02}
of the BdG Hamiltonian with a frequency-dependent order parameter.
This suggests a speculation that these finite-energy states could
potentially be used to reliably store quantum information.

\begin{figure}
\begin{centering}
\includegraphics[scale=0.22]{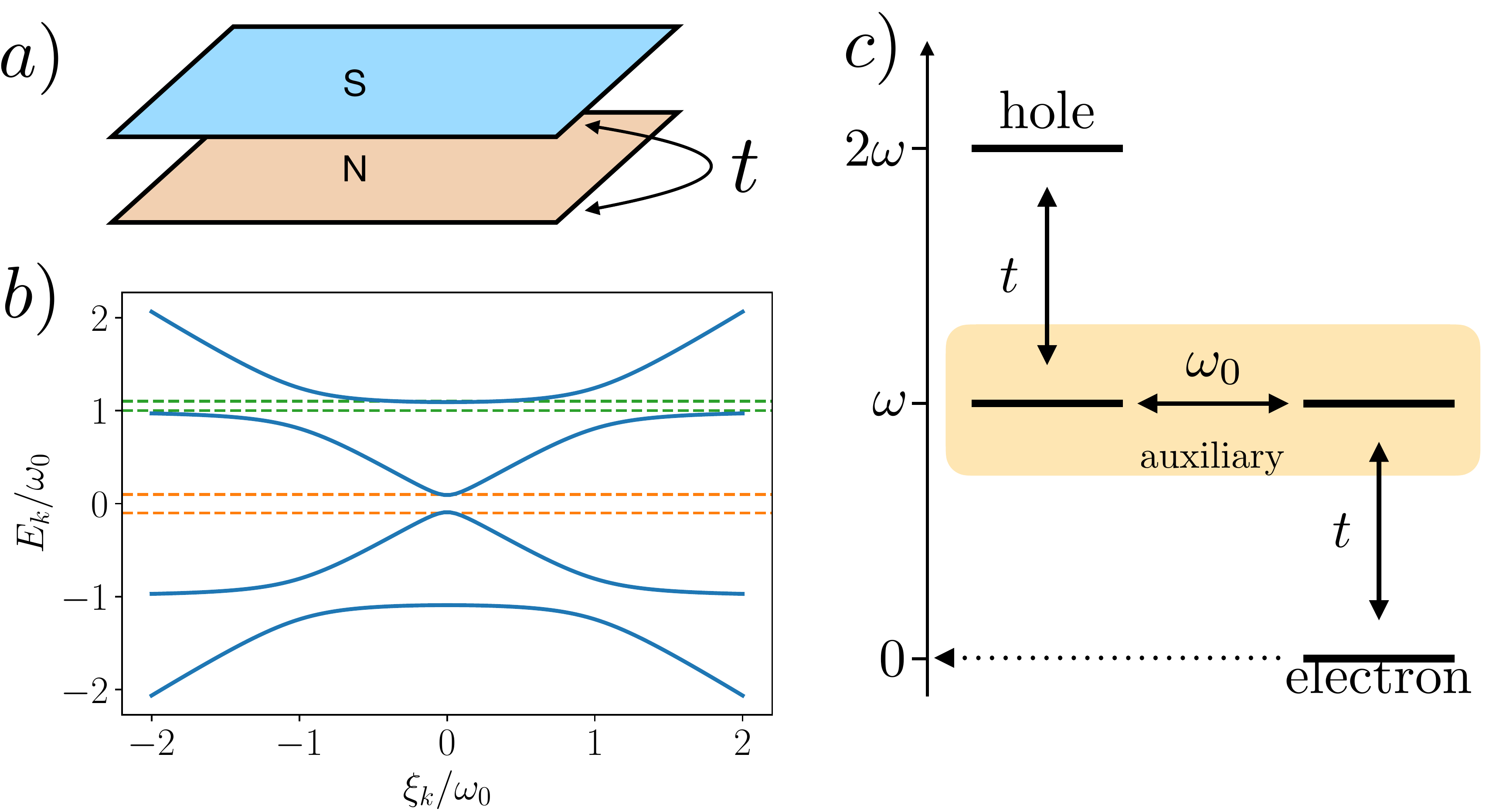} 
\par\end{centering}
\caption{a) Effective proximity system which induces a Lorentz-shape frequency
dependence of the order parameter. Fictitious flat-band superconductor
shown in blue and a normal-state metal is shown in orange. b) Quasiparticle
spectrum of the BdG equation with the frequency-dependent order parameter.
Additional band gaps are formed close to the characteristic frequency
of the order parameter frequency dependence. Blue solid lines correspond
to the eigenenergies of Eq.~(\ref{eq:G_tilde}). Orange dashed stands
for the conventional BCS gap ($\pm\phi_{0}$) and dashed green lines
$E_{k}=\omega_{0}$ and $E_{k}\approx\omega_{0}+\phi_{0}$ correspond
to the additional gap due to the frequency-dependence of the order
parameter. In the simulation we assumed $\phi_{0}=\omega_{0}/10$.
c) Schematic representation of the four-state system equivalent to
the on-shell quasi-electron branch of the BdG Hamiltonian Eq.~\eqref{eq:Hbdg}
in the limit of weak coupling $t\rightarrow0$. }

\label{Fig1} 
\end{figure}

\begin{figure}
\begin{centering}
\includegraphics[scale=0.55]{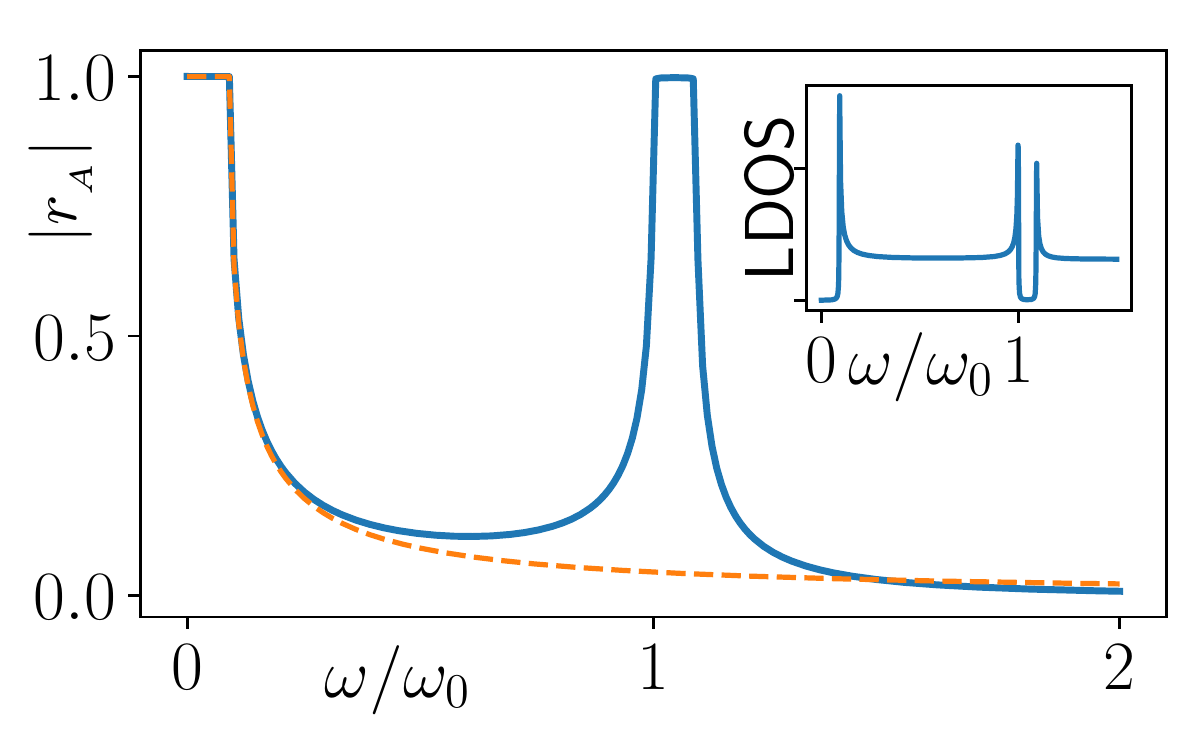} 
\par\end{centering}
\caption{Andreev reflection coefficient as function of the incident electron
energy. As result of the additional band gap the additional Andreev
reflection peak is observed at energies close to $\omega_{0}$. Blue
solid line corresponds to the BdG equation with the frequency-dependent
anomalous self-energy. Orange dashed stands for the conventional BdG
assuming no frequency dependence of the order parameter. Inset shows
the local density of states of the Green's function Eq.~\eqref{eq:SBDG}.
The gap is chosen such that $\phi_{0}=\omega_{0}/10$.}

\label{Fig2} 
\end{figure}

In this work we consider the conventional s-wave superconductor
but keeping a complete realistic frequency dependence of the order
parameter. We restrict our discussion to the mean-field level and
rely on the Bogoliubov-de Gennes approach, which requires a straightforward
generalization. Consider an electron gas with the creation
(annihilation) operators $\psi^{\dagger}_{{\bf k,\sigma=\shortuparrow,\downarrow}}$
$(\psi_{{\bf k},\sigma})$, where ${\bf k}$ denotes the
electron momentum. For the two-component spinor $\Psi_{{\bf k},n}=\{\psi_{{\bf k},\downarrow}\left(i\varepsilon_{n}\right),\psi_{-{\bf k},\uparrow}^{\dagger}\left(-i\varepsilon_{n}\right)\}^{T}$,
the Green function is defined as: $\hat{{\cal G}}_{{\bf k}}(i\varepsilon_{n})\equiv-\langle\Psi_{{\bf k},n}\otimes\Psi_{{\bf k},n}^{\dagger}\rangle$,
where $\varepsilon_{n}=\left(2n+1\right)\pi/\beta$ is the standard
the fermionic Matsubara frequency, $\beta$ is the inverse temperature,
and $n\in\mathbb{Z}$. Neglecting the momentum dependence of the self-energy,
the Green function of an interacting Fermi gas can be written as~\citep{M20, S18}:
\begin{equation}
\hat{{\cal G}}_{{\bf k}}^{-1}\left(i\varepsilon_{n}\right)=i\varepsilon_{n}\mathcal{Z}_{n}\hat{\tau}_{0}-\phi_{n}\hat{\tau}_{1}-\xi_{k}\hat{\tau}_{3},\label{eq:SBDG}
\end{equation}
where $\xi_{k}=k^{2}/2m-\mu$, $m$ is the electron mass, $\mu$ is
the chemical potential, $\tau_{i}$ are Pauli matrices. $\mathcal{Z}_{n}$
denotes the inverse quasiparticle residue obtained from the odd part of the
normal-state self-energy. The off-diagonal matrix element $\phi_{n}$
stands for the anomalous self energy. $\mathcal{Z}_{n}$ and $\phi_{n}$ naturally
appear in both intrinsic and proximity-induced superconductors \citep{CAE20,LSS19}.
The BdG equations correspond to $\hat{{\cal G}}_{{\bf k}}^{-1}\left(i\varepsilon\right)\chi=0$,
which parameterically determines the sought-after quasiparticle dispersion
(here, $\chi$ is a Nambu spinor).

To calculate the quasiparticle spectrum, we need an explicit form
of the frequency dependence of the anomalous self-energy. In what
follows, we consider the aforementioned Lorentz-shaped form (parameterized
by its amplitude $\phi_{0}$ and the characteristic frequency $\omega_{0}$)
\begin{equation}
\phi_{n}=\phi_{0}\,\frac{\omega_{0}^{2}}{\varepsilon_{n}^{2}+\omega_{0}^{2}}.\label{Lorentz}
\end{equation}
As demonstrated in Refs.~\citep{M18,MBM20} this solution naturally
appears in superconductors, where pairing is induced by an optical
phonon mode with the Einstein spectrum at weak coupling. We now consider
the spectrum of quasiparticles in Eq.~\eqref{eq:SBDG}.

To develop intuition, it is instructive to consider an auxiliary setup,
which involves a fictitious flat-band superconductor with a frequency-independent
gap proximity coupled to a normal metal as shown in Fig.~\ref{Fig1}. As we show, a proper choice
of parameters in this setup gives rise to a Green function, which
replicates Eq.~\ref{eq:SBDG} with the order parameter (\ref{Lorentz}).
The advantage of this construction is that its BdG Hamiltonian below
involves only standard, frequency-independent parameters: 
\begin{align}
\hat{{\cal G}}_{k}^{\prime-1}\left(i\varepsilon_{n}\right) & =i\varepsilon_{n}\hat{\tau}_{0}\hat{\sigma}_{0}-\hat{H}_{\text{BdG}}\label{eq:G_tilde}\\
H_{\text{BdG}}= & \xi_{k}\hat{\tau}_{3}\frac{\left(\hat{\sigma}_{3}+\hat{1}\right)}{2}-\omega_{0}\hat{\tau}_{1}\frac{\left(\hat{1}-\hat{\sigma}_{3}\right)}{2}+t\hat{\tau}_{3}\hat{\sigma}_{1},\label{eq:Hbdg}
\end{align}
Here $\hat{\tau}_{i}$ represents Pauli matrices in the Nambu space
and $\hat{\sigma}$ parametrizes an effective two-band model: $\left(\hat{\sigma}_{3}+\hat{1}\right)/2$
projects on the normal metal fermion modes and $\left(\hat{1}-\hat{\sigma}_{3}\right)/2$
projector on the fictitious flat-band superconductor. The order parameter
in the latter is set the characteristic phonon frequency, $\omega_{0}$.
We note that we assumed no intrinsic order parameter for the original
fermions. The coefficient $t$ denotes the tunneling amplitude between
the superconductor and the metal. The auxiliary superconductor can
be integrated-out generating both the anomalous and normal self energies.
By choosing $t=\sqrt{\phi_{0}\omega_{0}}$ we can exactly match frequency
dependence of the order parameter to reproduce Eq.~\ref{Lorentz}.
The corresponding $\mathcal{Z}$-factor in Eq.~(\ref{eq:SBDG}) is
equal to $\mathcal{Z}_{n}=\left[1+\phi_{0}\omega_{0}/(\omega_{0}^{2}+\varepsilon_{n}^{2})\right]\approx1$,
for $\phi_{0}\ll\omega_{0}$. This procedure effectively replaces
the integrating out the bosonic Einstein-phonon degree of freedom
(which generates the frequency dependence of the gap in the physical setup) with the integrating out the degrees of freedom of the auxiliary flat-band superconductor. Note that this picture is introduced for the purposes of illustration only. All results can reproduced in the original model directly. As we discuss in the supplementary material, the inverse quasiparicle residue, $\mathcal{Z}_n$ cannot be set
identically to one because it would result in an unstable spectrum.

The quasiparticle spectrum can now be found by diagonalizing the static
BdG Hamiltonian $H_{\text{BdG}}$ given in Eq.~\eqref{eq:Hbdg}.
The result is shown on Fig. ~\ref{Fig1} (b) and it has two band gaps.
First, we observe the conventional BCS-like band gap at low energies
$[-\phi_{0},\phi_{0}]$. It can be obtained by e.g. neglecting the
frequency-dependence of the self-energy in Eq.~\eqref{eq:SBDG},
which reduces it to the textbook case. The second band gap at high
energies is a specific feature of the two-band system as defined in
Eq.~\eqref{eq:G_tilde}. As a result of the flatness of one of the
dispersion relations, the avoided crossing forms a band gap close
to the frequency, $\omega_{0}$. The value of this second band gap
can be readily obtained analytically from Eq.~\eqref{eq:G_tilde}:
\begin{equation}
\phi_{2}=\sqrt{\frac{\omega_{0}}{2}\left(\sqrt{\omega_{0}(4\phi_{0}+\omega_{0})}+\phi_{0}+\frac{\omega_{0}}{2}\right)}-\omega_{0}\approx\phi_{0},\label{eq:gap}
\end{equation}
where the approximate sign corresponds to the limit $\phi_{0}\ll\omega_{0}$,
which must hold for weak coupling.

We now define the local density of states (LDOS) of the electron gas
as $\frac{-1}{\pi}\int d\xi_{k}\text{Im}\left(G_{{\bf k}}\left(\omega+i0^{+}\right)\right)_{1,1}$,
where $G_{{\bf k}}$ is the Green function~\ref{eq:G_tilde} analytically
continued to real frequencies. As the direct consequence of the band
gap the LDOS is strongly depleted at energies $\approx[\omega_{0},\omega_{0}+\phi_{0}]$
as shown on inset in Fig.~\ref{Fig2}. We note that the exactly zero
density of states is a feature of the flat-band dispersion of the
auxiliary superconductor/Einstein phonons. However, as we discuss
in the SM, introduction of a finite curvature to the phonon dispersion
would still lead to a significant depletion of the density of states.
As we discuss below this secondary gap can host additional Andreev
\citep{SJ18} reflection peaks, observable in metal-superconductor
heterostructures.

We now explore how the additional sharp Lorentz-like features of the
gap function affect the superconducting proximity effect. In order
to describe the transmission and reflection of quasiparticles, we
employ the Blonder-Tinkham-Klapwijk (BTK) formalism~\citep{BTK82}.
We consider a heterostructure consisting of normal and superconducting
metals (NS). Following \citep{BTK82},
we consider the scattering of an incident electron off of the barrier.
The strength of the proximity effect can be characterized by the probability
for an electron to scatter into a hole-type excitation.

Within the BTK theory the boundary condition is given by: 
\begin{align}
\vec{\psi}_{N}\left(0\right) & =\vec{\psi}_{S}\left(0\right),\label{eq:BC_1}\\
\frac{\partial_{z}}{2m_{N}}\vec{\psi}_{N}\left(0\right) & =\frac{\partial_{z}}{2m_{S}}\vec{\psi}_{S}\left(0\right)+H\vec{\psi}_{S}(0).\label{eq:BC_2}
\end{align}
where $m_{S}$ and $m_{N}$ are the effective electron masses and
$H$ is the $\delta$-barrier height. 
Performing the analytic continuation and replacing $i\epsilon_{n}\rightarrow\omega+i0^{+}$,
the wavefunctions on superconducting side satisfy the equation $\hat{G}_{{\bf k}}^{-1}(\omega+i0^{+})\vec{\psi}_{S}=0$.
On the normal side the equation is the same with the substitution
$\phi(\omega+i0^{+})\rightarrow0$ and $\mathcal{Z}(\omega+i0^{+})\rightarrow1$.
In the following we do not explicitly write $0^{+}$ for shortness.
The normal-state solution representing an incident electron and reflected
electron and hole components is:

\begin{equation}
\vec{\psi}_{N}=\begin{bmatrix}1\\
0
\end{bmatrix}e^{ik_{\text{e}}z}+\begin{bmatrix}r_{N}\\
0
\end{bmatrix}e^{-ik_{\text{e}}z}+\begin{bmatrix}0\\
r_{A}
\end{bmatrix}e^{ik_{\text{h}}z},\label{eq:psi_N}
\end{equation}
where $r_{N}$ and $r_{A}$ denote the reflection amplitude in the
electron and hole channels respectively and the electron/hole momenta
are given by $k_{\text{e}/\text{h}}=\sqrt{2m\left(\mu\pm\omega\right)}$.
Analogously we find the solution for the quasi-electrons and quasi-holes
propagating in the superconductor: 

\begin{align}
\vec{\psi}_{S} & \approx C_{\text{qe}}\begin{bmatrix}1\\
\eta_{+}
\end{bmatrix}e^{ik_{\text{qe}}z}+C_{\text{qh}}\begin{bmatrix}1\\
\eta_{-}
\end{bmatrix}e^{-ik_{\text{qh}}z},\label{eq:Psi_S}
\end{align}
where $C_{\text{qe}}$ and $C_{\text{qh}}$ are the corresponding
amplitudes of the quasi-electron and quasi-holes and we denoted the
corresponding coherence factors as $\eta_{\pm}=\phi(\omega)/\left(\mathcal{Z}(\omega)\omega\pm\sqrt{\mathcal{Z}^{2}(\omega)\omega^{2}-\phi^{2}(\omega)}\right)$.
In the quasiclassical limit the quasielectron and quasihole momenta
$k_{\text{qe}},k_{\text{qh}}$ can be taken to be equal to the corresponding
Fermi momenta.

Upon solving the set of equations Eqs.~(\ref{eq:BC_1}-\ref{eq:Psi_S})
in the quasi-classical \cite{BWF99} limit and assuming $m_{S}=m_{N}$
we find the Andreev reflection coefficient to be:

\begin{align}
r_{\text{A}} & =\frac{\eta_{-}\eta_{+}}{\eta_{-}+Z^{2}\left(\eta_{-}-\eta_{+}\right)},\label{eq:t_A}\\
r_{N} & =\frac{-Z\left(i+Z\right)}{\eta_{-}+Z^{2}\left(\eta_{-}-\eta_{+}\right)}\left(\eta_{-}-\eta_{+}\right)\label{eq:t_N}
\end{align}
where the normalized barrier height is defined as $Z\equiv mH/\sqrt{2m\mu}$.
We note that the conventional Andreev reflection can be obtained from
Eq.~(\ref{eq:t_A}) by simply assuming frequency-independent $\mathcal{Z}$
and $\phi$. In the limit of $\mathcal{Z}=0$ the Andreev reflection coefficient
is given by $r_{A}=\eta_{+}$. We therefore find that in order to
have a strong Andreev reflection the condition $\phi\gg \mathcal{Z}\omega$
should be satisfied. The latter condition is always satisfied at very
low frequencies leading to the conventional Andreev reflection \citep{SJ18}
result $\left|r_{\text{A}}\right|\approx1$. However it can also be
satisfied at large frequencies if the frequency-dependent order parameter
$\Delta\left(\omega\right)\equiv\phi\left(\omega\right)/\mathcal{Z}\left(\omega\right)$
is larger than the frequency $\Delta\left(\omega\right)\gg\omega$.
In this case the reflection is up to a possible phase factor identical
to the low-energy case. As shown in Fig.~\ref{Fig2}, this scenario is realized for the Lorentz-like order parameter introduced above in the frequency range close to the
resonance $\omega\sim\omega_{0}$. 
\paragraph{Finite-energy bound states --}

We now consider the possibility of having the finite-energy Andreev
bound states \citep{SJ18} in the junction consisting of two superconductors separated
by a metallic region with the two boundaries located at $L/2$ and $-L/2$. We assume the phase difference
between two superconductors to be $\gamma$. In order to find the
bound states we now follow the same procedure as for outlined above
for the Andreev reflection but matching solutions at the two boundaries simultaneously.
The general solution can be obtained analytically but it is too cumbersome
and we therefore consider some simple limiting case. In particular
as we demonstrate in the supplementary material, in the limit $\phi_{0}\ll\omega_{0}$
and $L\rightarrow0$ there are two additional in-gap bound states
$\omega\in[\omega_{0},\omega_{0}+\phi_{0}]$ with the energies $\omega_{\alpha}$
(both positive and negative):

\begin{equation}
\omega_{\alpha}=\omega_{0}+\frac{1}{2}\left\{ 1+\alpha\cos\frac{\gamma}{2}\right\} ,\label{eq:w_a}
\end{equation}
with $\alpha=\pm$. We thus find the energy of these bound states
is of the order of the characteristic frequency of the order parameter
frequency dependence. For both intrinsic and proximity superconductors
$\omega_{0}$ can be expected to be of the order of several THz. This
implies the existence of such bound states can potentially be probed by means
of laser excitation. Their response is equivalent to a two-level system
thus making them a good candidate for realization of solid state qubits.

\paragraph*{Interpretation as \textquotedbl dark\textquotedbl{} resonance --}

We now discuss the interpretation of the additional Andreev reflection
peak in terms of the so-called ``dark'' resonance. The concept of
dark resonance is extensively studied within the field of quantum
optics \citep{LYF99}. It is based on the existence of ``slowly''-evolving
superposition states in a quantum system which are decoupled from
the ``fast'' e.g. environment modes. For example, such optical phenomena
as the Electromagnetically Induced Transparency (EIT) \citep{FL00}
are based on the existence of a dark resonance in a driven three-level
system. The on-shell BdG Green's function Eq.~\eqref{eq:G_tilde}, can be expressed as follows:
\[
\hat{G}_{{\bf k}\rightarrow{\bf k}(\omega)}^{\prime}\left(\omega+i0^{+}\right)=\left(\begin{array}{cccc}
\omega-\xi_{\pm}\left(\omega\right) & 0 & t & 0\\
0 & \omega+\xi_{\pm}\left(\omega\right) & 0 & -t\\
t & 0 & \omega & \omega_{0}\\
0 & -t & \omega_{0} & \omega
\end{array}\right)
\]
where $\xi_{\pm}\left(\omega\right)=\pm\sqrt{\left(\left(t^{2}-\omega^{2}\right)^{2}-\omega^{2}\omega_{0}^{2}\right)/(\omega^{2}-\omega_{0}^{2})}$
is the on-shell  quasi-electron and quasi-hole dispersions.
By construction the Green function has the flat-band superconducting degree
of freedom, which can be considered ``slow.'' By ``dark'' we thus define states, which  have projection
onto the auxiliary degrees of freedom only. Let us now find the quasi-electron
and quasi-hole coherence vectors corresponding to the on-shell Green's
function Eq.~(\ref{eq:G_tilde}, \ref{eq:Hbdg}). The latter is schematically
shown in Fig.~\ref{Fig1}~(c) in the limit of $t\rightarrow0$.
At frequencies $\omega\approx\omega_{0}$ one of the eigenstates of
the auxiliary degrees of freedom crosses $\omega=0$ therefore
being degenerate with the electron branch. The corresponding eigenvector
is readily found to be given by $\approx[0,0,1,1]^{T}/\sqrt{2}$.
Thus this vector only has \textquotedbl slow\textquotedbl{} non-propagating
(flat-band) components and are decoupled from the other degrees of
freedom.

\begin{center}
\begin{figure}
\begin{centering}
\includegraphics[scale=0.6]{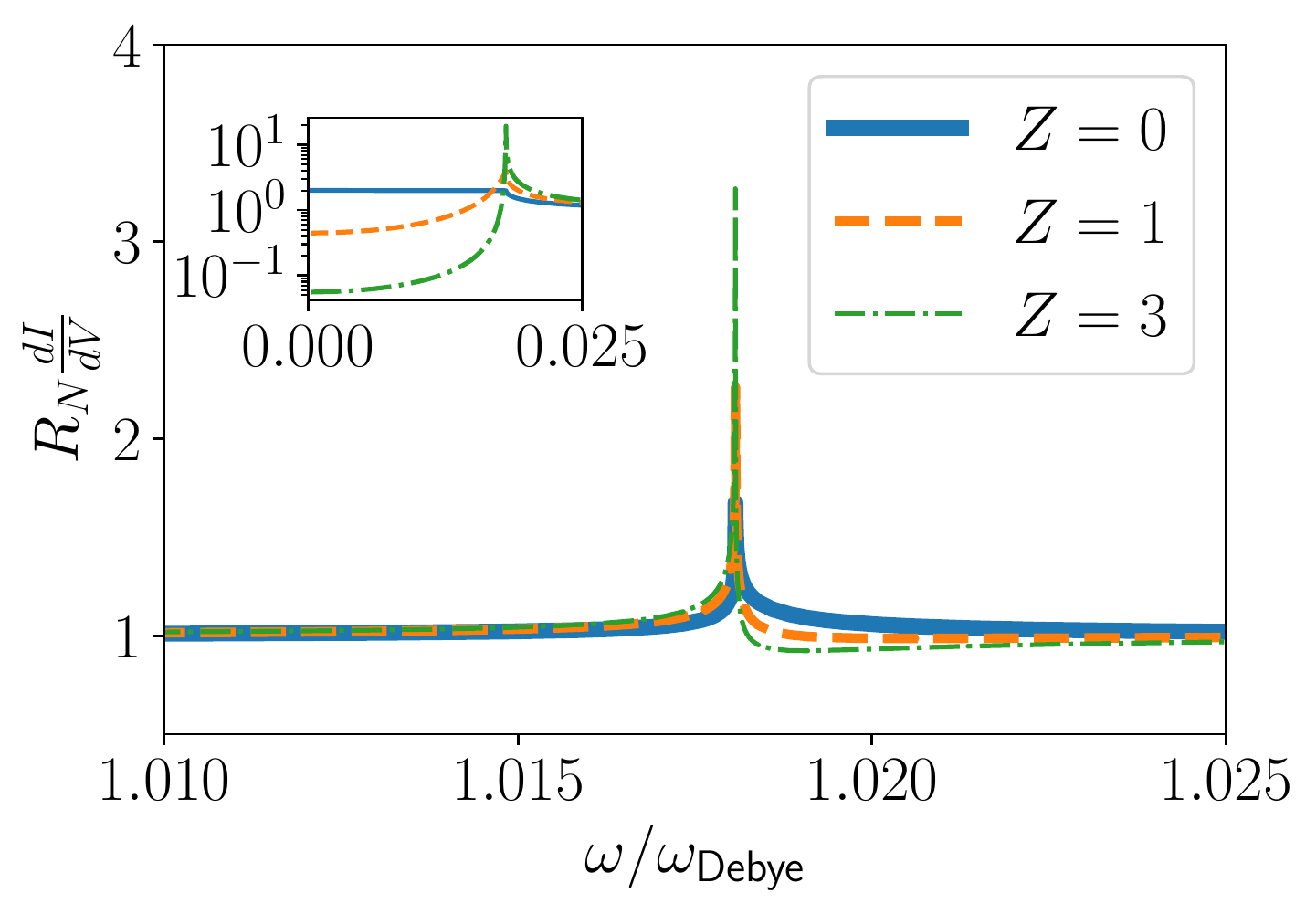}
\par\end{centering}
\caption{Differential tunneling conductance as function of bias voltage (expressed
in frequency units) for different values of the barrier height: $Z=0$
solid blue, $Z=1$ dashed orange and $Z=3$ dot-dashed green. Inset
shows the same at low frequencies. The assumed electron-phonon coupling
is $\lambda=0.3$ (see supplementary material). The normal-state resistance is denoted as $R_N=(Z^2+1)/(2\nu_0 e^2 v_F \mathcal {A})$, where $\mathcal{A}$ is the contact surface area. }
\label{Fig_Phonon}
\end{figure}
\par\end{center}

\paragraph{Einstein phonon model --}
 Finally we demonstrate that key results and conclusions of the toy model, involving the auxiliary superconductor, hold in the physical setup of interest, where  the pairing interaction is induced by the optical phonon mode. More specifically we now consider the gap function
induced by the interaction with the optical phonon mode with the propagator
$\mathcal{D_{{\bf {q}}}}(i\nu_{m})=-2\omega_{0}/(\omega_{0}^{2}+\nu_{m}^{2}),$
where $\nu_{m}\equiv2\pi m/\beta$,
$m\in\mathbb{Z}$. As we demonstrate in the Supplementary material,
the matrix-valued self-energy $\hat{\Sigma}(i\epsilon_{n})$ is found
by solving the Dyson's equation. The latter reduces to two coupled
equations for the inverse quasiparticle residue $\mathcal{Z}_{n}$ and  the
gap function $\phi_{n}$ in case when the phonon propagator is momentum-independent:
$\hat{\Sigma}(i\epsilon_{n})=(1-\mathcal{Z}_{n})i\epsilon_{n}\hat{\tau}_{0}+\phi_{n}\hat{\tau}_{1}$.
The approximate analytical form of the phonon-induced self-energy
can be obtained in the limit of weak electron-phonon coupling.
As was shown in \citep{M18,MBM20}, the imaginary-axis dependence of the order parameter has the Lorentz form $\phi_{n}\propto\omega_{0}^{2}/(\omega_{0}^{2}+\epsilon_{n}^{2})$
and ${\cal Z}_{n}\approx1$. This is thus in agreement with the assumed
gap-frequency behavior in Eq.~\eqref{eq:SBDG}. At finite electron-phonon
coupling strength, the gap frequency dependence deviates from the Lorentz form and additional resonances at frequencies $n\omega_{0}$
with $n =1, 2, 3 \ldots$ \citep{MBM20} emerge. However, the sharp features of the gap function, reminiscent to the pure Lorentz case, remain even at finite
coupling strength \cite{M20,MBM20,M18}. We now study how these features
affect the Andreev reflection from the NS boundary. More precisely,
we consider the differential tunneling conductance which expresses
through the reflection coefficients Eqs.~(\ref{eq:t_A}, \ref{eq:t_N})
as $dI/dV\propto1+|r_{A}|^{2}-|r_{N}|^{2}$ \cite{BTK82}. Both reflection
coefficients are found numerically by solving the complete set
of Migdal-Eliashberg equations. The result of numerical calculation is shown in Fig.~\ref{Fig_Phonon} for different values of the barrier height $Z$. We find the prominent additional reflection peaks close to the Debye frequency, which correspond to the dark Andreev resonances. Note that the exact solution for the phonon case involves imaginary self-energy contributions, which give rise to  a finite life-time of the superconducting quasiparticles (absent in the toy model). However, these complications do not appear to affect the qualitative picture and the signatures of the dark Andreev states are preserved.

\paragraph{Conclusions \& outlook --} In this work, we studied the quasiparticle  properties of a superconductor with a frequency-dependent order parameter. When the latter has resonant  features, we find an additional depletion of the high-energy density of states. We provide a physically equivalent  two-band picture with an additional  superconducting band gap emerging at high energies. For the NS junction, the band gap leads to additional Andreev reflection peaks at high energies. In the case of an SNS junction, we find Andreev bound states within the high-energy band gap.  We provide an interpretation of these phenomena in terms of the ``dark'' resonance of the BdG Hamiltonian. We expect that the predicted phenomena should be accessible in experiment  in  superconductors with optical-phonon-mediated pairing (for example in $MgB_2$ \cite{MAJ02}, $K_3 C_{60}$ \cite{ZOR91}, etc) and also proximity systems involving flat-band materials \cite{BDE20}. This work also suggests a number of follow up ideas, at the intersection of superconductivity and quantum optics: e.g., the possibility of control of the dark Andreev states by external laser driving. Furthermore, it would be interesting to explore the role of frequency dependence of the order parameter on the quasiparticle spectrum in topological superconductors and whether dark Majorana-like states are possible.

\paragraph{Acknowledgements --} This work was supported by the National Science Foundation under Grant No. DMR-2037158, the U.S. Army Research Office under Contract No. W911NF1310172, and the Simons Foundation. The authors are grateful to Jay Sau for an illuminating discussion.

 \bibliographystyle{apsrev4-1}
\bibliography{bibl}

\begin{thebibliography}{20}%
\makeatletter
\providecommand \@ifxundefined [1]{%
 \@ifx{#1\undefined}
}%
\providecommand \@ifnum [1]{%
 \ifnum #1\expandafter \@firstoftwo
 \else \expandafter \@secondoftwo
 \fi
}%
\providecommand \@ifx [1]{%
 \ifx #1\expandafter \@firstoftwo
 \else \expandafter \@secondoftwo
 \fi
}%
\providecommand \natexlab [1]{#1}%
\providecommand \enquote  [1]{``#1''}%
\providecommand \bibnamefont  [1]{#1}%
\providecommand \bibfnamefont [1]{#1}%
\providecommand \citenamefont [1]{#1}%
\providecommand \href@noop [0]{\@secondoftwo}%
\providecommand \href [0]{\begingroup \@sanitize@url \@href}%
\providecommand \@href[1]{\@@startlink{#1}\@@href}%
\providecommand \@@href[1]{\endgroup#1\@@endlink}%
\providecommand \@sanitize@url [0]{\catcode `\\12\catcode `\$12\catcode
  `\&12\catcode `\#12\catcode `\^12\catcode `\_12\catcode `\%12\relax}%
\providecommand \@@startlink[1]{}%
\providecommand \@@endlink[0]{}%
\providecommand \url  [0]{\begingroup\@sanitize@url \@url }%
\providecommand \@url [1]{\endgroup\@href {#1}{\urlprefix }}%
\providecommand \urlprefix  [0]{URL }%
\providecommand \Eprint [0]{\href }%
\providecommand \doibase [0]{http://dx.doi.org/}%
\providecommand \selectlanguage [0]{\@gobble}%
\providecommand \bibinfo  [0]{\@secondoftwo}%
\providecommand \bibfield  [0]{\@secondoftwo}%
\providecommand \translation [1]{[#1]}%
\providecommand \BibitemOpen [0]{}%
\providecommand \bibitemStop [0]{}%
\providecommand \bibitemNoStop [0]{.\EOS\space}%
\providecommand \EOS [0]{\spacefactor3000\relax}%
\providecommand \BibitemShut  [1]{\csname bibitem#1\endcsname}%
\let\auto@bib@innerbib\@empty
\bibitem [{\citenamefont {Bardeen}\ \emph {et~al.}(1957)\citenamefont
  {Bardeen}, \citenamefont {Cooper},\ and\ \citenamefont {Schrieffer}}]{BCS57}%
  \BibitemOpen
  \bibfield  {author} {\bibinfo {author} {\bibfnamefont {J.}~\bibnamefont
  {Bardeen}}, \bibinfo {author} {\bibfnamefont {L.~N.}\ \bibnamefont {Cooper}},
  \ and\ \bibinfo {author} {\bibfnamefont {J.~R.}\ \bibnamefont {Schrieffer}},\
  }\href {\doibase 10.1103/PhysRev.108.1175} {\bibfield  {journal} {\bibinfo
  {journal} {Phys. Rev.}\ }\textbf {\bibinfo {volume} {108}},\ \bibinfo {pages}
  {1175} (\bibinfo {year} {1957})}\BibitemShut {NoStop}%
\bibitem [{\citenamefont {Altland}\ and\ \citenamefont {Simons}(2010)}]{AS10}%
  \BibitemOpen
  \bibfield  {author} {\bibinfo {author} {\bibfnamefont {A.}~\bibnamefont
  {Altland}}\ and\ \bibinfo {author} {\bibfnamefont {B.~D.}\ \bibnamefont
  {Simons}},\ }\href@noop {} {\emph {\bibinfo {title} {Condensed matter field
  theory}}}\ (\bibinfo  {publisher} {Cambridge university press},\ \bibinfo
  {year} {2010})\BibitemShut {NoStop}%
\bibitem [{\citenamefont {Eliashberg}(1960)}]{E60}%
  \BibitemOpen
  \bibfield  {author} {\bibinfo {author} {\bibfnamefont {G.}~\bibnamefont
  {Eliashberg}},\ }\href@noop {} {\bibfield  {journal} {\bibinfo  {journal}
  {Sov. Phys. JETP}\ }\textbf {\bibinfo {volume} {11}},\ \bibinfo {pages} {696}
  (\bibinfo {year} {1960})}\BibitemShut {NoStop}%
\bibitem [{\citenamefont {Marsiglio}(2020)}]{M20}%
  \BibitemOpen
  \bibfield  {author} {\bibinfo {author} {\bibfnamefont {F.}~\bibnamefont
  {Marsiglio}},\ }\href {\doibase https://doi.org/10.1016/j.aop.2020.168102}
  {\bibfield  {journal} {\bibinfo  {journal} {Annals of Physics}\ }\textbf
  {\bibinfo {volume} {417}},\ \bibinfo {pages} {168102} (\bibinfo {year}
  {2020})},\ \bibinfo {note} {eliashberg theory at 60: Strong-coupling
  superconductivity and beyond}\BibitemShut {NoStop}%
\bibitem [{\citenamefont {Marsiglio}\ \emph {et~al.}(1988)\citenamefont
  {Marsiglio}, \citenamefont {Schossmann},\ and\ \citenamefont
  {Carbotte}}]{MSC88}%
  \BibitemOpen
  \bibfield  {author} {\bibinfo {author} {\bibfnamefont {F.}~\bibnamefont
  {Marsiglio}}, \bibinfo {author} {\bibfnamefont {M.}~\bibnamefont
  {Schossmann}}, \ and\ \bibinfo {author} {\bibfnamefont {J.~P.}\ \bibnamefont
  {Carbotte}},\ }\href {\doibase 10.1103/PhysRevB.37.4965} {\bibfield
  {journal} {\bibinfo  {journal} {Phys. Rev. B}\ }\textbf {\bibinfo {volume}
  {37}},\ \bibinfo {pages} {4965} (\bibinfo {year} {1988})}\BibitemShut
  {NoStop}%
\bibitem [{\citenamefont {Marsiglio}(2018)}]{M18}%
  \BibitemOpen
  \bibfield  {author} {\bibinfo {author} {\bibfnamefont {F.}~\bibnamefont
  {Marsiglio}},\ }\href@noop {} {\bibfield  {journal} {\bibinfo  {journal}
  {Physical Review B}\ }\textbf {\bibinfo {volume} {98}},\ \bibinfo {pages}
  {024523} (\bibinfo {year} {2018})}\BibitemShut {NoStop}%
\bibitem [{\citenamefont {Mirabi}\ \emph {et~al.}(2020)\citenamefont {Mirabi},
  \citenamefont {Boyack},\ and\ \citenamefont {Marsiglio}}]{MBM20}%
  \BibitemOpen
  \bibfield  {author} {\bibinfo {author} {\bibfnamefont {S.}~\bibnamefont
  {Mirabi}}, \bibinfo {author} {\bibfnamefont {R.}~\bibnamefont {Boyack}}, \
  and\ \bibinfo {author} {\bibfnamefont {F.}~\bibnamefont {Marsiglio}},\
  }\href@noop {} {\bibfield  {journal} {\bibinfo  {journal} {Physical Review
  B}\ }\textbf {\bibinfo {volume} {101}},\ \bibinfo {pages} {064506} (\bibinfo
  {year} {2020})}\BibitemShut {NoStop}%
\bibitem [{\citenamefont {Christensen}\ and\ \citenamefont
  {Chubukov}(2021)}]{CC21}%
  \BibitemOpen
  \bibfield  {author} {\bibinfo {author} {\bibfnamefont {M.~H.}\ \bibnamefont
  {Christensen}}\ and\ \bibinfo {author} {\bibfnamefont {A.~V.}\ \bibnamefont
  {Chubukov}},\ }\href@noop {} {\bibfield  {journal} {\bibinfo  {journal}
  {Physical Review B}\ }\textbf {\bibinfo {volume} {104}},\ \bibinfo {pages}
  {L140501} (\bibinfo {year} {2021})}\BibitemShut {NoStop}%
\bibitem [{\citenamefont {Fleischhauer}\ and\ \citenamefont
  {Lukin}(2000)}]{FL00}%
  \BibitemOpen
  \bibfield  {author} {\bibinfo {author} {\bibfnamefont {M.}~\bibnamefont
  {Fleischhauer}}\ and\ \bibinfo {author} {\bibfnamefont {M.~D.}\ \bibnamefont
  {Lukin}},\ }\href {\doibase 10.1103/PhysRevLett.84.5094} {\bibfield
  {journal} {\bibinfo  {journal} {Phys. Rev. Lett.}\ }\textbf {\bibinfo
  {volume} {84}},\ \bibinfo {pages} {5094} (\bibinfo {year}
  {2000})}\BibitemShut {NoStop}%
\bibitem [{\citenamefont {Lukin}\ \emph {et~al.}(1999)\citenamefont {Lukin},
  \citenamefont {Yelin}, \citenamefont {Fleischhauer},\ and\ \citenamefont
  {Scully}}]{LYF99}%
  \BibitemOpen
  \bibfield  {author} {\bibinfo {author} {\bibfnamefont {M.~D.}\ \bibnamefont
  {Lukin}}, \bibinfo {author} {\bibfnamefont {S.~F.}\ \bibnamefont {Yelin}},
  \bibinfo {author} {\bibfnamefont {M.}~\bibnamefont {Fleischhauer}}, \ and\
  \bibinfo {author} {\bibfnamefont {M.~O.}\ \bibnamefont {Scully}},\ }\href
  {\doibase 10.1103/PhysRevA.60.3225} {\bibfield  {journal} {\bibinfo
  {journal} {Phys. Rev. A}\ }\textbf {\bibinfo {volume} {60}},\ \bibinfo
  {pages} {3225} (\bibinfo {year} {1999})}\BibitemShut {NoStop}%
\bibitem [{\citenamefont {Fleischhauer}\ and\ \citenamefont
  {Lukin}(2002)}]{FM02}%
  \BibitemOpen
  \bibfield  {author} {\bibinfo {author} {\bibfnamefont {M.}~\bibnamefont
  {Fleischhauer}}\ and\ \bibinfo {author} {\bibfnamefont {M.~D.}\ \bibnamefont
  {Lukin}},\ }\href@noop {} {\bibfield  {journal} {\bibinfo  {journal}
  {Physical Review A}\ }\textbf {\bibinfo {volume} {65}},\ \bibinfo {pages}
  {022314} (\bibinfo {year} {2002})}\BibitemShut {NoStop}%
\bibitem [{\citenamefont {Schrieffer}(2018)}]{S18}%
  \BibitemOpen
  \bibfield  {author} {\bibinfo {author} {\bibfnamefont {J.~R.}\ \bibnamefont
  {Schrieffer}},\ }\href@noop {} {\emph {\bibinfo {title} {Theory of
  superconductivity}}}\ (\bibinfo  {publisher} {CRC press},\ \bibinfo {year}
  {2018})\BibitemShut {NoStop}%
\bibitem [{\citenamefont {Chubukov}\ \emph {et~al.}(2020)\citenamefont
  {Chubukov}, \citenamefont {Abanov}, \citenamefont {Esterlis},\ and\
  \citenamefont {Kivelson}}]{CAE20}%
  \BibitemOpen
  \bibfield  {author} {\bibinfo {author} {\bibfnamefont {A.~V.}\ \bibnamefont
  {Chubukov}}, \bibinfo {author} {\bibfnamefont {A.}~\bibnamefont {Abanov}},
  \bibinfo {author} {\bibfnamefont {I.}~\bibnamefont {Esterlis}}, \ and\
  \bibinfo {author} {\bibfnamefont {S.~A.}\ \bibnamefont {Kivelson}},\
  }\href@noop {} {\bibfield  {journal} {\bibinfo  {journal} {Annals of
  Physics}\ }\textbf {\bibinfo {volume} {417}},\ \bibinfo {pages} {168190}
  (\bibinfo {year} {2020})}\BibitemShut {NoStop}%
\bibitem [{\citenamefont {Liu}\ \emph {et~al.}(2019)\citenamefont {Liu},
  \citenamefont {Sau}, \citenamefont {Stanescu},\ and\ \citenamefont
  {Das~Sarma}}]{LSS19}%
  \BibitemOpen
  \bibfield  {author} {\bibinfo {author} {\bibfnamefont {C.-X.}\ \bibnamefont
  {Liu}}, \bibinfo {author} {\bibfnamefont {J.~D.}\ \bibnamefont {Sau}},
  \bibinfo {author} {\bibfnamefont {T.~D.}\ \bibnamefont {Stanescu}}, \ and\
  \bibinfo {author} {\bibfnamefont {S.}~\bibnamefont {Das~Sarma}},\ }\href
  {\doibase 10.1103/PhysRevB.99.024510} {\bibfield  {journal} {\bibinfo
  {journal} {Phys. Rev. B}\ }\textbf {\bibinfo {volume} {99}},\ \bibinfo
  {pages} {024510} (\bibinfo {year} {2019})}\BibitemShut {NoStop}%
\bibitem [{\citenamefont {Sauls}(2018)}]{SJ18}%
  \BibitemOpen
  \bibfield  {author} {\bibinfo {author} {\bibfnamefont {J.}~\bibnamefont
  {Sauls}},\ }\href@noop {} {\enquote {\bibinfo {title} {Andreev bound states
  and their signatures},}\ } (\bibinfo {year} {2018})\BibitemShut {NoStop}%
\bibitem [{\citenamefont {Blonder}\ \emph {et~al.}(1982)\citenamefont
  {Blonder}, \citenamefont {Tinkham},\ and\ \citenamefont {Klapwijk}}]{BTK82}%
  \BibitemOpen
  \bibfield  {author} {\bibinfo {author} {\bibfnamefont {G.~E.}\ \bibnamefont
  {Blonder}}, \bibinfo {author} {\bibfnamefont {M.}~\bibnamefont {Tinkham}}, \
  and\ \bibinfo {author} {\bibfnamefont {T.~M.}\ \bibnamefont {Klapwijk}},\
  }\href {\doibase 10.1103/PhysRevB.25.4515} {\bibfield  {journal} {\bibinfo
  {journal} {Phys. Rev. B}\ }\textbf {\bibinfo {volume} {25}},\ \bibinfo
  {pages} {4515} (\bibinfo {year} {1982})}\BibitemShut {NoStop}%
\bibitem [{\citenamefont {Belzig}\ \emph {et~al.}(1999)\citenamefont {Belzig},
  \citenamefont {Wilhelm}, \citenamefont {Bruder}, \citenamefont {Sch{\"o}n},\
  and\ \citenamefont {Zaikin}}]{BWF99}%
  \BibitemOpen
  \bibfield  {author} {\bibinfo {author} {\bibfnamefont {W.}~\bibnamefont
  {Belzig}}, \bibinfo {author} {\bibfnamefont {F.~K.}\ \bibnamefont {Wilhelm}},
  \bibinfo {author} {\bibfnamefont {C.}~\bibnamefont {Bruder}}, \bibinfo
  {author} {\bibfnamefont {G.}~\bibnamefont {Sch{\"o}n}}, \ and\ \bibinfo
  {author} {\bibfnamefont {A.~D.}\ \bibnamefont {Zaikin}},\ }\href@noop {}
  {\bibfield  {journal} {\bibinfo  {journal} {Superlattices and
  microstructures}\ }\textbf {\bibinfo {volume} {25}},\ \bibinfo {pages} {1251}
  (\bibinfo {year} {1999})}\BibitemShut {NoStop}%
\bibitem [{\citenamefont {Mazin}\ \emph {et~al.}(2002)\citenamefont {Mazin},
  \citenamefont {Andersen}, \citenamefont {Jepsen}, \citenamefont {Dolgov},
  \citenamefont {Kortus}, \citenamefont {Golubov}, \citenamefont {Kuz'menko},\
  and\ \citenamefont {Van Der~Marel}}]{MAJ02}%
  \BibitemOpen
  \bibfield  {author} {\bibinfo {author} {\bibfnamefont {I.}~\bibnamefont
  {Mazin}}, \bibinfo {author} {\bibfnamefont {O.}~\bibnamefont {Andersen}},
  \bibinfo {author} {\bibfnamefont {O.}~\bibnamefont {Jepsen}}, \bibinfo
  {author} {\bibfnamefont {O.}~\bibnamefont {Dolgov}}, \bibinfo {author}
  {\bibfnamefont {J.}~\bibnamefont {Kortus}}, \bibinfo {author} {\bibfnamefont
  {A.~A.}\ \bibnamefont {Golubov}}, \bibinfo {author} {\bibfnamefont
  {A.}~\bibnamefont {Kuz'menko}}, \ and\ \bibinfo {author} {\bibfnamefont
  {D.}~\bibnamefont {Van Der~Marel}},\ }\href@noop {} {\bibfield  {journal}
  {\bibinfo  {journal} {Physical review letters}\ }\textbf {\bibinfo {volume}
  {89}},\ \bibinfo {pages} {107002} (\bibinfo {year} {2002})}\BibitemShut
  {NoStop}%
\bibitem [{\citenamefont {Zhang}\ \emph {et~al.}(1991)\citenamefont {Zhang},
  \citenamefont {Ogata},\ and\ \citenamefont {Rice}}]{ZOR91}%
  \BibitemOpen
  \bibfield  {author} {\bibinfo {author} {\bibfnamefont {F.~C.}\ \bibnamefont
  {Zhang}}, \bibinfo {author} {\bibfnamefont {M.}~\bibnamefont {Ogata}}, \ and\
  \bibinfo {author} {\bibfnamefont {T.~M.}\ \bibnamefont {Rice}},\ }\href
  {\doibase 10.1103/PhysRevLett.67.3452} {\bibfield  {journal} {\bibinfo
  {journal} {Phys. Rev. Lett.}\ }\textbf {\bibinfo {volume} {67}},\ \bibinfo
  {pages} {3452} (\bibinfo {year} {1991})}\BibitemShut {NoStop}%
\bibitem [{\citenamefont {Balents}\ \emph {et~al.}(2020)\citenamefont
  {Balents}, \citenamefont {Dean}, \citenamefont {Efetov},\ and\ \citenamefont
  {Young}}]{BDE20}%
  \BibitemOpen
  \bibfield  {author} {\bibinfo {author} {\bibfnamefont {L.}~\bibnamefont
  {Balents}}, \bibinfo {author} {\bibfnamefont {C.~R.}\ \bibnamefont {Dean}},
  \bibinfo {author} {\bibfnamefont {D.~K.}\ \bibnamefont {Efetov}}, \ and\
  \bibinfo {author} {\bibfnamefont {A.~F.}\ \bibnamefont {Young}},\ }\href@noop
  {} {\bibfield  {journal} {\bibinfo  {journal} {Nature Physics}\ }\textbf
  {\bibinfo {volume} {16}},\ \bibinfo {pages} {725} (\bibinfo {year}
  {2020})}\BibitemShut {NoStop}%
\end{thebibliography}%

\newpage
\newpage 

\onecolumngrid
\newpage
{
\center \bf \large 
Supplemental Material for: \\
Dark Andreev States in Superconductors\vspace*{0.1cm}\\ 
\vspace*{0.0cm}
}
\begin{center}
Andrey Grankin, Victor Galitski\\
\vspace*{0.15cm}
\small{\textit{Joint Quantum Institute, Department of Physics, University of Maryland, College Park, MD 20742, USA}}\\
\vspace*{0.25cm}
\end{center}

\twocolumngrid

\section{Derivation of the Green's function in the toy model}

In this section we trace-over the auxiliary fermionic degrees of freedom
and derive the Green's function of the superconductor. We start with
the BDG Hamiltonian:

\begin{align}
H_{\text{BdG}}= & \xi_{k}\hat{\tau}_{3}\frac{\left(\hat{\sigma}_{3}+\hat{1}\right)}{2}-\omega_{0}\hat{\tau}_{1}\frac{\left(\hat{1}-\hat{\sigma}_{3}\right)}{2}+t\hat{\tau}_{3}\hat{\sigma}_{1},\label{eq:Hbdg-1}
\end{align}
The quantum-mechanical action corresponding to Eq.~\eqref{eq:Hbdg-1}
reads:

\begin{equation}
S=-\sum_{n}\tilde{\Psi}_{{\bf k}}^{\dagger}{\cal G}_{{\bf k}}^{'-1}(i\epsilon_{n})\tilde{\Psi}_{{\bf k}},\label{eq:S}
\end{equation}
where $\tilde{\Psi}$ is the extended Bogolyubov spinor and ${\cal G}_{{\bf k}}^{\prime-1}(i\epsilon_{n})=i\epsilon_{n}-H_{\text{BdG}}$.
The partitioning function is related to Eq.~\eqref{eq:S} $Z=\int d\ldots e^{-S}$.
We now perform the Gaussian integral over the auxiliary degrees of
freedom parametrized by the projector $Q\equiv\left(\hat{1}-\hat{\sigma}_{3}\right)/2$.
The projector onto the remaining fermionic degrees of freedom reads
${\cal P}={\cal I}-{\cal Q}$. The Green's function is straightforwardly
found to be:

\begin{align*}
{\cal G}_{{\bf k}}^{-1}\left(i\epsilon_{n}\right) & =i\epsilon_{n}-\xi_{k}\hat{\tau}_{3}-t^{2}\hat{\tau}_{3}\frac{1}{i\epsilon_{n}+\omega_{0}\hat{\tau}_{1}}\hat{\tau}_{3}\\
 & =i\epsilon_{n}-\xi_{k}\hat{\tau}_{3}+t^{2}\hat{\tau}_{3}\frac{i\epsilon_{n}-\omega_{0}\hat{\tau}_{1}}{\epsilon_{n}^{2}+\omega_{0}^{2}}\hat{\tau}_{3}\\
 & =i\epsilon_{n}-\xi_{k}\hat{\tau}_{3}+t^{2}\frac{i\epsilon_{n}+\omega_{0}\tau_{1}}{\epsilon_{n}^{2}+\omega_{0}^{2}}
\end{align*}
We thus find the additional normal- and anomalous self-energies $-t^{2}i\epsilon_{n}/(\epsilon_{n}^{2}+\omega_{0}^{2}),$
$-t^{2}\omega_{0}/(\epsilon_{n}^{2}+\omega_{0}^{2})$. Using $t\equiv\sqrt{\phi_{0}\omega_{0}}$
as discussed in the main text we find 
\[
{\cal Z}_{n}=\left(1+\frac{\phi_{0}\omega_{0}}{\epsilon_{n}^{2}+\omega_{0}^{2}}\right).
\]

\subsection{Instability in the absence of ${\cal Z}_{n}$}

\begin{figure}
\begin{centering}
\includegraphics[scale=0.45]{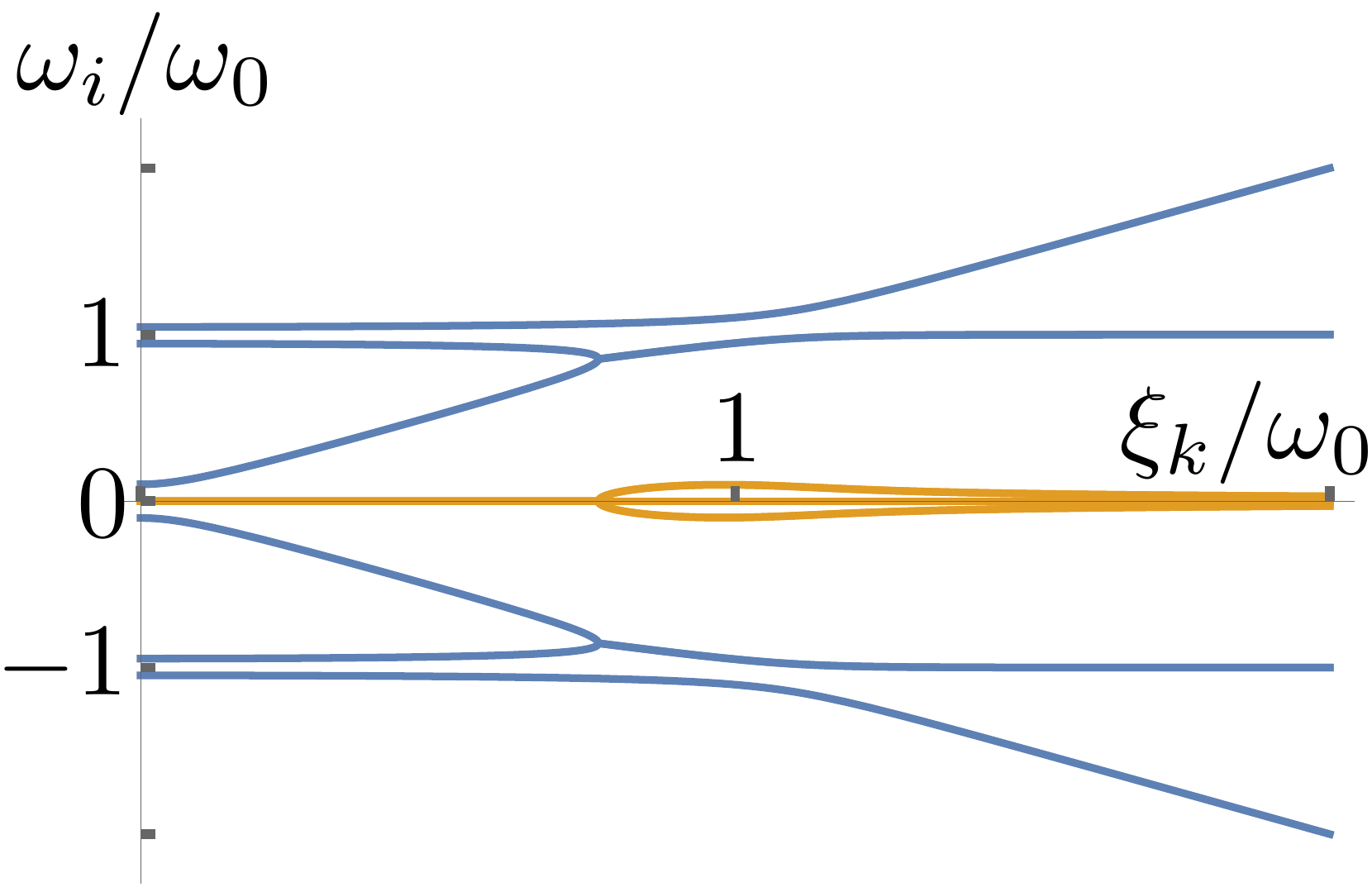} 
\par\end{centering}
\caption{Real (blue) and imaginary (orange) parts of the eigenvalues of the BdG Green's function Eq.~\eqref{eq:SM_unstable}. The non-vanishing imaginary parts correspond to the  eigenvalues merging close to the $\omega_0$ frequency.}
\label{FIG_SM}
\end{figure}

Let us now consider a superconductor with the frequency-dependent
order parameter and no renormalization of the normal-state self energy:

\begin{equation}
\hat{{\cal G}}_{{\bf k}}^{-1}\left(i\varepsilon_{n}\right)=i\varepsilon_{n}\hat{\tau}_{0}-\phi_{0}\frac{\omega_{0}^{2}}{\epsilon_{n}^{2}+\omega_{0}^{2}}\hat{\tau}_{1}-\xi_{k}\hat{\tau}_{3},\label{eq:SBDG-1}
\end{equation}
We now find the spectrum of this Greens function and demonstrate that
it is unstable.

\[
\hat{G}_{{\bf k}}^{-1}\left(\omega+i0^{+}\right)=\left(\omega+i0^{+}\right)\hat{\tau}_{0}-\phi_{0}\frac{\omega_{0}^{2}}{\omega_{0}^{2}-\left(\omega+i0^{+}\right)^{2}}\hat{\tau}_{1}-\xi_{k}\hat{\tau}_{3}\label{eq:SM_unstable}
\]

\[
\det\hat{G}_{{\bf k}}^{-1}\left(\omega+i0^{+}\right)=0
\]
The eigenvalues $\omega_i$ are shown in Fig.~\ref{FIG_SM}. We find that that the spectrum becomes
complex with both positive and negative imaginary values. This corresponds
to an unstable system. We thus conclude that the frequency-dependence of the gap implies a presence of a non-zero quasiparticle residue function $\mathcal{Z}_n$.

\section{Derivation of the gap function at weak coupling\label{sec:Derivation-of-the}}

Dyson equation for the matrix-valued self energy reads:

\begin{equation}
\hat{\Sigma}_{{\bf k}}(i\varepsilon_{n})=-\frac{g^{2}}{V\beta}\sum_{{\bf q},m}{\cal D}_{{\bf k-k'}}(i\varepsilon_{n}-i\epsilon_{n'})\left\{ \hat{\tau}_{3}\hat{G}_{{\bf k'}}(i\epsilon_{n'})\hat{\tau}_{3}\right\} \label{eq:Sigma}
\end{equation}
The propagator of the Einstein phonon mode reads

\[
{\cal D}_{{\bf k-k'}}(i\varepsilon_{n}-i\epsilon_{n'})=\frac{-2\omega_{0}}{\omega_{0}^{2}-(i\varepsilon_{n}-i\epsilon_{n'})^{2}},
\]
where $\omega_{0}$ is the Debye frequency. Taking into account the
fact that the phonon propagator does not depend on momentum we can
simplify Eq.~(\ref{eq:Sigma}):

\begin{equation}
\hat{\Sigma}(i\varepsilon_{n})=-\frac{g^{2}\nu_{0}}{\beta}\sum_{m}{\cal D}(i\varepsilon_{n}-i\epsilon_{n'})\int_{-\infty}^{\infty}d\xi_{k'}\left\{ \hat{\tau}_{3}\hat{G}_{{\bf k'}}(i\epsilon_{n'})\hat{\tau}_{3}\right\} .\label{eq:Sigma-1}
\end{equation}
The momentum integral $\int d\xi_{k'}$ can be taken analytically
by employing the Green's function ansatz Eq.~(\ref{eq:SBDG}).

\subsubsection{Weak-coupling result \label{subsec:Weak-coupling-result}}

Let us now explicitly find the gap function. Let us first write the
complete set of Eliashberg equations as obtained from Eq.~(\ref{eq:Sigma-1}):

\begin{align}
Z_{n} & =1+\frac{\lambda}{\beta}\sum_{m}\frac{\omega_{0}^{2}}{\omega_{0}^{2}+\left(\varepsilon_{n}-\epsilon_{n'}\right)^{2}}\frac{\pi\frac{\epsilon_{n'}}{\varepsilon_{n}}Z_{m}}{\sqrt{\epsilon_{n'}^{2}Z_{m}^{2}+\phi_{m}^{2}}}\\
\Delta_{n}Z_{n} & =\frac{\lambda}{\beta}\sum_{m}\frac{\omega_{0}^{2}}{\omega_{0}^{2}+\left(\varepsilon_{n}-\epsilon_{n'}\right)^{2}}\frac{\pi\Delta_{m}}{\sqrt{\epsilon_{n'}^{2}+\Delta_{m}^{2}}}.\label{eq:DeltaZ}
\end{align}
where we defined $\Delta_{m}\equiv\phi_{m}/Z_{m}$. At very weak coupling
we can neglect the non-linearity of both equations or equivalently
set $\sqrt{\epsilon_{n'}^{2}+\Delta_{m}^{2}}\approx\left|\epsilon_{n'}\right|$.
In this case Eq.~(\ref{eq:DeltaZ}) has logarithmic singularity on
the righthand side. Following \citep{M18} we can separate singular
and regular terms in a straightforward way:

\begin{align*}
\Delta_{n}Z_{n} & =\frac{\lambda}{\beta}\sum_{m}\left(\frac{\omega_{0}^{2}}{\omega_{0}^{2}+\left(\varepsilon_{n}-\epsilon_{n'}\right)^{2}}-\frac{\omega_{0}^{2}}{\omega_{0}^{2}+\left(\varepsilon_{n}\right)^{2}}\right)\frac{\pi\Delta_{m}}{\left|\epsilon_{n'}\right|}\\
 & +\frac{\lambda}{\beta}\sum_{m}\left(\frac{\omega_{0}^{2}}{\omega_{0}^{2}+\left(\varepsilon_{n}\right)^{2}}\right)\frac{\pi\Delta_{m}}{\left|\epsilon_{n'}\right|}
\end{align*}
The first term to the righthand side is now regular while the second
one is singular. It is therefore leading at very weak coupling $\lambda\rightarrow0$.
In this case we get:

\[
\Delta_{n}\propto\frac{\omega_{0}^{2}}{\omega_{0}^{2}+\left(\varepsilon_{n}\right)^{2}}
\]
We note that in the limit $\lambda\rightarrow0$ the anomalous self-energy
has the same functional form\cite{MBM20,M18}. Analytical continuation
requires some caution and the result will be shown in the section
below. 

\subsection{Analytic continuation}

Following the formulation derived in \cite{M18,MBM20,MSC88} we now
perform the so-called iterative analytic continuation procedure of
the self-energy Eq.~(\ref{eq:Sigma}). Let us now use the spectral
decomposition of fermionic and bosonic 
\begin{align}
{\cal D}\left(i\varepsilon_{n}-i\epsilon_{n'}\right) & =\frac{1}{\pi}\int dx\frac{\text{Im}\left\{ D^{R}\left(x\right)\right\} }{x-(i\varepsilon_{n}-i\epsilon_{n'})},\\
\hat{{\cal G}}_{{\bf k}}(i\epsilon_{n'}) & =\frac{1}{\pi}\int dy\frac{\text{Im}\left\{ \hat{G}_{{\bf k}}^{R}\left(y\right)\right\} }{y-i\epsilon_{n'}}.
\end{align}
where $D^{R}$ and $\hat{G}^{R}$ stand for the retarded Green's functions.
By substituting these expressions into Eq.~(\ref{eq:Sigma}) and
taking explicitly the Matsubara-frequency sum over $\epsilon_{n'}$
we find:

\begin{align*}
 & \hat{\Sigma}(i\varepsilon_{n})=\\
 & -g^{2}\nu_{0}\int_{-\infty}^{\infty}\frac{d\xi_{k}}{\pi^{2}}\int dxdy\frac{\text{Im}\left\{ D^{R}\left(x\right)\right\} \text{Im}\left\{ \hat{\tau}_{3}\hat{G}_{{\bf k}}^{R}\left(y\right)\hat{\tau}_{3}\right\} }{y-i\varepsilon_{n}-x}\left(n_{x}+f_{y}\right),
\end{align*}
where $n_{x}$ and $f_{y}$ are respectively bosonic and fermionic
occupation numbers. Analytic continuation is readily performed by
replacing $i\varepsilon_{n}\rightarrow\omega+i0^{+}$.

\begin{align*}
 & \hat{\Sigma}\left(\omega+i0^{+}\right)=\\
 & -g^{2}\nu_{0}\int_{-\infty}^{\infty}\frac{d\xi_{k}}{\pi^{2}}\int dxdy\frac{\text{Im}\left\{ D^{R}\left(x\right)\right\} \text{Im}\left\{ \hat{\tau}_{3}\hat{G}_{{\bf k}}^{R}\left(y\right)\hat{\tau}_{3}\right\} }{y-x-\omega-i0^{+}}\left(n_{x}+f_{y}\right)
\end{align*}
The integral over $y$ can now be taken explicitly. For that we use
the following facts: retarded Green's function only has poles in the
lower part of the complex plane.

\begin{align*}
 & \hat{\Sigma}\left(\omega+i0^{+}\right)=\\
 & -g^{2}\nu_{0}\int_{-\infty}^{\infty}\frac{d\xi_{k}}{\pi}\int dx\text{Im}\left\{ D^{R}\left(x\right)\right\} \left\{ \hat{\tau}_{3}\hat{G}_{{\bf k}}^{R}\left(x+\omega\right)\hat{\tau}_{3}\right\} \left(n_{x}+f_{x+\omega}\right)\\
 & -\frac{g^{2}\nu_{0}}{\beta}\sum_{n}\int_{-\infty}^{\infty}d\xi_{k}{\cal D}\left(\omega-i\varepsilon_{n}\right)\hat{\tau}_{3}\hat{{\cal G}}_{{\bf k}}\left(i\varepsilon_{n}\right)\hat{\tau}_{3}
\end{align*}
The right-hand side of this expression contains both the Matsubara
and the real-frequency Green's function. We can now estimate the gap-frequency
function in the limit of very weak coupling. We will be interested
in the following quantity (see main text) $\Delta\left(\omega+i0^{+}\right)\equiv\phi\left(\omega+i0^{+}\right)/Z\left(\omega+i0^{+}\right)$.

Let us now derive the real-axis $\lambda\rightarrow0$ expression
for the anomalous self-energy. By taking into account $\text{Im}\left\{ D^{R}\left(x\right)\right\} =\pi\left(\delta\left(\omega+\omega_{0}\right)-\delta\left(\omega-\omega_{0}\right)\right)$
at low we find:

\begin{align*}
 & \hat{\Sigma}\left(\omega+i0^{+}\right)=\\
 & +g^{2}\nu_{0}\int_{-\infty}^{\infty}d\xi_{k}\left\{ \hat{\tau}_{3}\hat{G}_{{\bf k}}^{R}\left(\omega_{0}+\omega\right)\hat{\tau}_{3}\right\} \left(n_{\omega_{0}}+f_{\omega_{0}+\omega}\right)\\
 & -g^{2}\nu_{0}\int_{-\infty}^{\infty}d\xi_{k}\left\{ \hat{\tau}_{3}\hat{G}_{{\bf k}}^{R}\left(\omega-\omega_{0}\right)\hat{\tau}_{3}\right\} \left(n_{-\omega_{0}}+f_{-\omega_{0}+\omega}\right)\\
 & -\frac{g^{2}\nu_{0}}{\beta}\sum_{n}\int_{-\infty}^{\infty}d\xi{\cal D}\left(\omega-i\varepsilon_{n}\right)\hat{\tau}_{3}\hat{{\cal G}}_{{\bf k}}\left(i\varepsilon_{n}\right)\hat{\tau}_{3}
\end{align*}
The first term is exponentially suppressed at $T\ll\omega_{0}$. The
second term is non-singular and moreover in the $T\rightarrow0$ limit
it only has contributes at $\omega>\omega_{0}$. The logarithmically-divergent
contribution in the $\lambda\rightarrow0$ is given by the second
term. In this limit the anomalous self-energy obeys:

\begin{align*}
\phi(\omega+i0^{+}) & \approx-\frac{g^{2}\nu_{0}}{\beta}\sum_{n}\int_{-\infty}^{\infty}d\xi_{k}{\cal D}\left(\omega-i\varepsilon_{n}\right)\frac{\phi_{n}}{\varepsilon_{n}^{2}\mathcal{Z}_{n}^{2}+\phi_{n}^{2}+\xi_{k}^{2}}\\
 & =\lambda\pi\int\frac{d\omega_{n}}{2\pi}\frac{\omega_{0}^{2}}{\omega_{0}^{2}-(\omega-i\epsilon_{n})^{2}}\frac{\phi_{n}}{\sqrt{\varepsilon_{n}^{2}\mathcal{Z}_{n}^{2}+\phi_{n}^{2}}}
\end{align*}
Under the same assumptions as in Sec.~\eqref{subsec:Weak-coupling-result}
we find:

\[
\phi(\omega+i0^{+})\propto\frac{\omega_{0}^{2}}{\omega_{0}^{2}-(\omega+i0^{+})^{2}}.
\]

\subsection{Normal state self-energy in the weak-coupling limit}

We now consider the diagonal terms of the normal-state self-energy
in the limit of weak coupling.

\subsubsection{Real-frequency}

\begin{align*}
 & \Sigma\left(\omega+i0^{+}\right)=\\
 & -2\pi ig^{2}\nu_{0}\theta\left(\omega-\omega_{0}\right)-ig^{2}\nu_{0}\int d\varepsilon_{n}\frac{2\omega_{0}\text{sign}(\epsilon_{n})}{\omega_{0}^{2}-\left(\omega-i\varepsilon_{n}\right)^{2}}\\
 & =-2\pi ig^{2}\nu_{0}\theta\left(\omega-\omega_{0}\right)-ig^{2}\nu_{0}\left(-4i\text{arctanh}\left(\frac{\omega}{\omega_{0}}\right)+2\pi\theta\left(\omega-\omega_{0}\right)\right)\\
 & =-\pi i\lambda\omega_{0}\theta\left(\omega-\omega_{0}\right)-\lambda\omega_{0}\text{arctanh}\left(\frac{\omega}{\omega_{0}}\right)
\end{align*}

\section{High-energy Andreev Bound states}

We now derive the energy of Andreev bound states appearing at high
energy due to the additional band gap. Our consideration is essentially
very similar to the derivation of the Andreev reflection coefficient
in the main text. In particular we now respectively denote the wave
functions of the right and left superconductors as $\vec{\psi}_{S,R}$
and $\vec{\psi}_{S,L}$. Solving the equation of motion in all three
media we find 
\begin{widetext}
:

\begin{align}
\vec{\psi}_{\text{S,R}} & =\begin{bmatrix}1\\
\frac{e^{i\gamma}\phi(\omega)}{\mathcal{Z}(\omega)\omega+\sqrt{Z^{2}(\omega)\omega^{2}-\phi^{2}(\omega)}}
\end{bmatrix}C_{\text{qe}}^{R}e^{ik_{\text{qe}}z}+\begin{bmatrix}1\\
\frac{e^{i\gamma}\phi(\omega)}{\mathcal{Z}(\omega)\omega-\sqrt{Z^{2}(\omega)\omega^{2}-\phi^{2}(\omega)}}
\end{bmatrix}C_{\text{qh}}^{R}e^{-ik_{\text{qh}}z},\label{eq:psiR}\\
\vec{\psi}_{\text{S,L}} & =\begin{bmatrix}1\\
\frac{\phi(\omega)}{\mathcal{Z}(\omega)\omega+\sqrt{Z^{2}(\omega)\omega^{2}-\phi^{2}(\omega)}}
\end{bmatrix}C_{\text{qe}}^{L}e^{-ik_{\text{qe}}z}+\begin{bmatrix}1\\
\frac{\phi(\omega)}{\mathcal{Z}(\omega)\omega-\sqrt{Z^{2}(\omega)\omega^{2}-\phi^{2}(\omega)}}
\end{bmatrix}C_{\text{qh}}^{L}e^{ik_{\text{qh}}z},\\
\vec{\psi}_{\text{N}} & =\begin{bmatrix}\alpha_{N}^{+}\\
0
\end{bmatrix}e^{ik_{\text{e}}z}+\begin{bmatrix}\alpha_{N}^{-}\\
0
\end{bmatrix}e^{-ik_{\text{e}}z}+\begin{bmatrix}0\\
\beta^{+}
\end{bmatrix}e^{-ik_{\text{h}}z}+\begin{bmatrix}0\\
\beta^{-}
\end{bmatrix}e^{ik_{\text{h}}z}.
\end{align}
\end{widetext}

In Eq.~(\ref{eq:psiR}) we introduced an additional global phase
factor $e^{i\gamma}$ due to the possible global phase difference
between the two superconductors. The problem can be solved numerically
or in the limit when the thickness of the normal metal is very thin
$L\rightarrow0$. In this case we just match boundaries between two
superconductors:

\begin{align}
\vec{\psi}_{\text{S,L}}\left(0\right) & \approx\vec{\psi}_{\text{S,R}}\left(0\right)\\
\partial_{z}\vec{\psi}_{\text{S,L}}\left(0\right) & \approx\partial_{z}\psi_{\text{S,R}}\left(0\right)
\end{align}
These two equations define the spectrum of the bound states. The bound
state energy $\omega_{\alpha}$ is readily obtained from the equation:

\[
\phi^{2}\left(\omega_{\alpha}\right)-Z\left(\omega_{\alpha}\right)^{2}\omega_{\alpha}^{2}+\phi^{2}(\omega_{\alpha})\cos\gamma=0
\]
By using the Lorentz-form expressions of $\phi$ and $\mathcal{Z}$ in the limit
$\phi_{0}\ll\omega_{0}$ we find the bound state energies Eq.~(9)
of the main text. 

\section{Spectral function}

We now study the spectral function of the Green's function. It is
defined as:

\section{Density of states calculation}

In this section we consider the density of states of the superconductor
with the BdG Green's function

\begin{equation}
\hat{G}_{k}^{-1}\left(\omega+i0^{+}\right)=(\omega+i0^{+}){\cal I}_{4}-\begin{Bmatrix}\xi_{k} & 0 & t & 0\\
0 & -\xi_{k} & 0 & -t\\
t & 0 & \xi_{k}/\alpha & \omega_{0}\\
0 & -t & \omega_{0} & -\xi_{k}/\alpha
\end{Bmatrix},\label{eq:G_tilde-1}
\end{equation}

\begin{figure}
\begin{centering}
\includegraphics[scale=0.4]{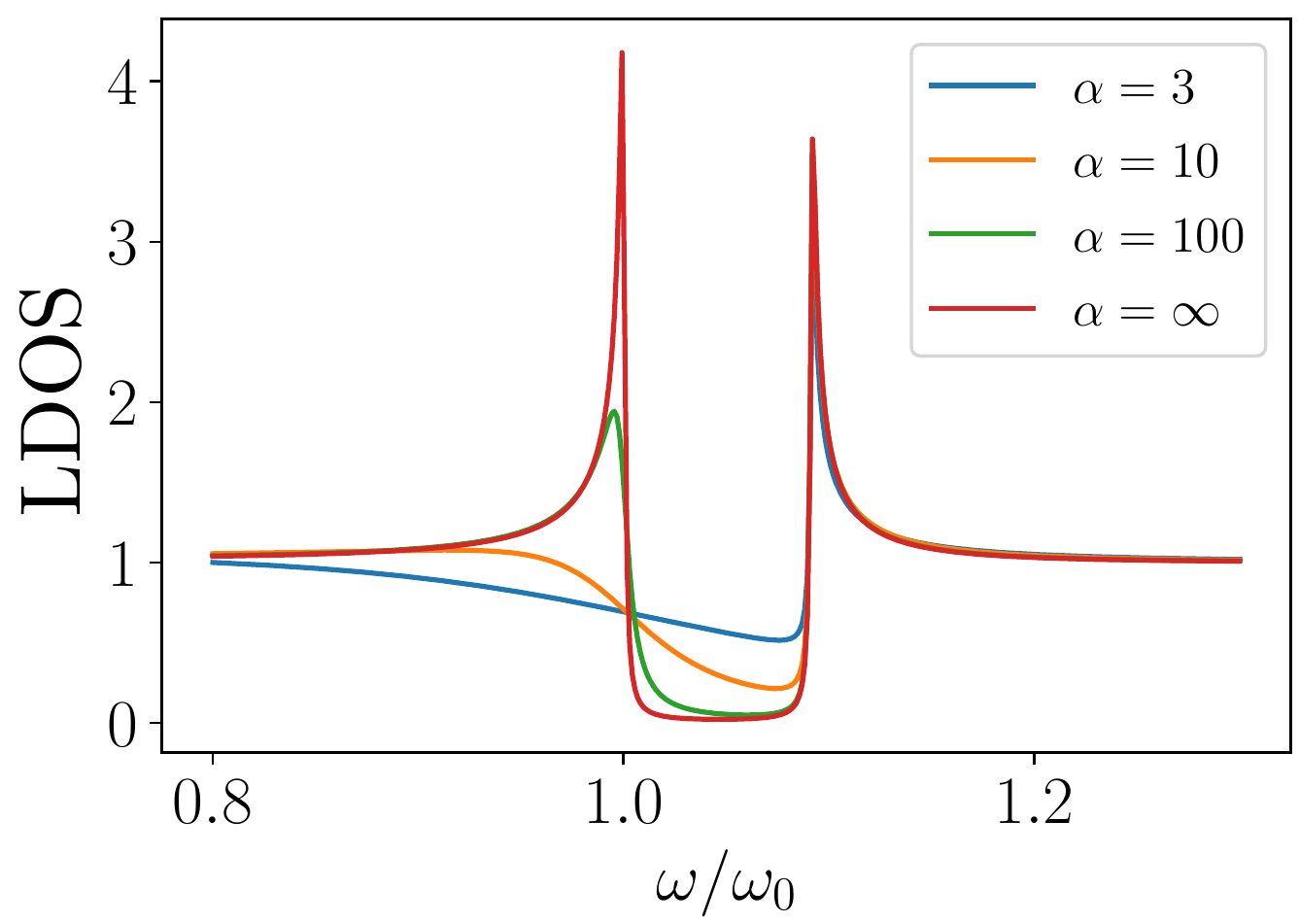} 
\par\end{centering}
\caption{Local density of states in Eq.~\ref{eq:G_tilde-1} as function of
the mass ratio parameter $\alpha$.}

\label{SM1} 
\end{figure}

This Green's function is almost the same as Eq.~\ref{eq:G_tilde}
in the main text except for instead of the flat-band superconductor
we take one with the dispersion $\xi_{k}/\alpha$ where $\alpha$
denotes the mass ratio parameter. The evolution of density of states
as function of $\alpha$ is shown in Fig.~\ref{SM1}. We therefore
observe a significant density of states depletion for $\alpha\gg3$.

\end{document}